\documentclass[12pt]{article}
\usepackage{epsf}
\usepackage[utf8]{inputenc}
\usepackage[english]{babel}

\begin{document}

\title{
{\bf The ground state of the Pomeron and its decays to light mesons and photons}}
\author{A.A. Godizov\thanks{E-mail: anton.godizov@gmail.com}\\
{\small {\it SRC Institute for High Energy Physics of NRC ``Kurchatov Institute'', 142281 Protvino, Russia}}}
\date{}
\maketitle

\begin{abstract}
The problem of the timelike Pomeron coupling to light mesons and photons is considered in light of available data on high-energy meson-proton scattering. 
Possible correspondence of $f_2(1950)$ resonance to the ground state of the Pomeron is argued.
\end{abstract}

\section*{1. Introduction}

In Regge phenomenology \cite{collins}, the Pomeron is introduced as an even Reggeon with vacuum quantum numbers and the intercept of its Regge trajectory higher than unity. 
Any Reggeon is defined as an analytic continuation of some set of bound states and resonances to the region of complex values of the spin. Many meson Reggeons emerging in 
the Quark Model have already got their identification via association with experimentally observed hadron states \cite{pdgf2}:\linebreak $f$-Reggeon ($f_2(1270)$ and 
$f_4(2050)$ mesons), $\rho$-Reggeon ($\rho(770)$ and $\rho_3(1690)$ mesons),\linebreak $a$-Reggeon, $\omega$-Reggeon, $\phi$-Reggeon, $\pi$-Reggeon, {\it etc.} All meson 
Regge trajectories have intercepts noticeably lower than unity. As a consequence, high-energy diffractive scattering of hadrons is dominated by Pomeron interaction. In 
particular, the representation of elastic scattering amplitude in terms of Pomeron exchanges yields a satisfactory description of the nucleon-nucleon diffractive pattern at 
ultrahigh collision energies and low momentum transfers \cite{godizov}.\linebreak However, being one of the most important objects of strong interaction physics (the main 
carrier of strong interaction in the high-energy diffraction domain), the Pomeron has not been discovered yet. The question {\it ``which of the observed $0^+(2^{++})$ states, 
if any, is the ground state of the Pomeron?''} still waits for proper answer. 

From the practical standpoint, true identification of the Pomeron is possible only through estimation of the partial decay widths and branching ratios of its real states 
(even-spin resonances). Presumably, the Pomeron resonance decays to pairs ``hadron-antihadron'' are driven by the same functions as the Pomeron coupling to the corresponding 
hadrons in high-energy diffractive scattering (see the Appendix). If we knew the analytic structure of these functions, we would be able to calculate the partial decay widths 
of the Pomeron real states. Unfortunately, at its present stage of development, QCD does not help to solve the problem. Nevertheless, some relevant information can be 
extracted from the available experimental data on meson-proton scattering. In this eprint, we will make an attempt to identify the ground state of the Pomeron, searching for 
interrelation between the high-energy hadron diffraction data and the measured characteristics of observed $0^+(2^{++})$ resonances.

For example, in the framework of the DL model \cite{dl}, the ratio of the kaon-proton and pion-proton total cross-sections is approximately equal (up to secondary Reggeon 
contributions) to the ratio of the Pomeron couplings to the corresponding mesons in forward scattering. At $\sqrt{s}=24.1$ GeV (the highest reached energy for $K p$ 
scattering), the impact of secondary Reggeons on the quantity \cite{pdg} 
\begin{equation}
\label{dlrat}
\frac{\sigma^{K^+p}_{tot}}{\sigma^{\pi^+p}_{tot}}\approx 0.844
\end{equation}
is minimal. As the partial decay width of the Pomeron ground state to pair of scalar mesons is proportional to the squared absolute value of the 
corresponding meson-Pomeron coupling, so, using (\ref{dlrat}), we obtain a qualitative estimation
\begin{equation}
\label{pomrat}
\frac{\Gamma_{{\rm P}\to K^+K^-}}{\Gamma_{{\rm P}\to \pi^+\pi^-}}\sim 0.844^2\approx 0.71\,.
\end{equation}

This value is consistent with the $f_2(1950)$ resonance branching ratio 
\begin{equation}
\label{f2rat}
\frac{\Gamma_{f_2(1950)\to K^+K^-}}{\Gamma_{f_2(1950)\to \pi^+\pi^-}}=0.565^{+0.19}_{-0.27}
\end{equation}
extracted from the the Belle Collaboration data on the quantities $\Gamma(\gamma\gamma)\Gamma(\pi^0\pi^0)/\Gamma_{tot}$ \cite{pion} and 
$\Gamma(\gamma\gamma)\Gamma(K^+K^-)/\Gamma_{tot}$ \cite{kaon}. On the one hand, such a proportion between the partial decay widths could be considered characteristic for any 
$SU(3)_f$ singlet. On the other hand, in the $t-J$ plane, $f_2(1950)$ resonance is located very close to the DL Pomeron Regge trajectory,\linebreak 
$\alpha_{\rm P}^{DL}(t)=1.08\,+\,0.25\,t$ \cite{dl2}, widely used in phenomenology. Therefore, it may be considered as a promising candidate for the Pomeron ground state. 

However, $0^+(2^{++})$ hadrons different from $f_2(1950)$ should not be immediately excluded from the consideration. Of course, the observed smallness of the branching ratios 
$\frac{\Gamma_{f_2(1270)\to K\bar K}}{\Gamma_{f_2(1270)\to \pi\pi}}$ and $\frac{\Gamma_{f'_2(1525)\to \pi\pi}}{\Gamma_{f'_2(1525)\to K\bar K}}$ \cite{pdgf2} allows to 
consider resonances $f_2(1270)$ and $f'_2(1525)$ as mesons of certain quark-antiquark content and to relate them to secondary Reggeons, but the current interpretation of 
other well-established $0^+(2^{++})$ resonances (namely, $f_2(2010)$, $f_2(2300)$, and $f_2(2340)$ \cite{pdgf2}) is not so clear. To distinguish reliably the Pomeron spin-2 
state among various vacuum resonances of the same spin, we need more information on the Pomeron dynamics in both the spacelike and timelike domains. First, let us estimate 
the Pomeron coupling to light mesons in the diffractive scattering regime.

\section*{2. The spacelike Pomeron coupling to light mesons}

\subsection*{2.1. Elastic scattering of scalar mesons on protons}

In the leading approximation, the angular distributions in $\pi^+p$ and $K^+p$ elastic scattering can be calculated in the same way as for $pp$ scattering (see \cite{godizov} 
and references therein):
$$
\frac{d\sigma}{dt} = \frac{|T(s,t)|^2}{16\pi s^2}\,\;,\;\;\;\;T(s,t) = 4\pi s\int_0^{\infty}db^2\,J_0(b\sqrt{-t})\,\frac{e^{2i\delta(s,b)}-1}{2i}\;,
$$
\begin{equation}
\label{eikrepr}
\delta(s,b)=\frac{1}{16\pi s}\int_0^{\infty}d(-t)\,J_0(b\sqrt{-t})\,\delta_{\rm P}(s,t)\approx
\end{equation}
$$
\approx\frac{1}{16\pi s}\int_0^{\infty}d(-t)\,J_0(b\sqrt{-t})\;
g_{hh\rm P}(t)\,g_{pp\rm P}(t)\left(i+{\rm tan}\frac{\pi(\alpha_{\rm P}(t)-1)}{2}\right)\pi\alpha'_{\rm P}(t)\left(\frac{s}{2s_0}\right)^{\alpha_{\rm P}(t)},
$$
where $s$ and $t$ are the Mandelstam variables, $b$ is the impact parameter, $s_0 = 1$ GeV$^2$, $\alpha_{\rm P}(t)$ is the Regge trajectory of the Pomeron, $g_{pp\rm P}(t)$ 
is the Pomeron coupling to proton, and $g_{hh\rm P}(t)$ is the Pomeron coupling to the incoming meson. At $t<0$, the Pomeron Regge trajectory and the Pomeron coupling to 
proton are approximated by the same test functions as in nucleon-nucleon scattering \cite{godizov}, 
\begin{equation}
\label{pomeron}
\alpha_{\rm P}(t) = 1+\frac{\alpha_{\rm P}(0)-1}{1-\frac{t}{\tau_a}}\;,\;\;\;\;g_{pp\rm P}(t)=\frac{g_{pp\rm P}(0)}{(1-a_gt)^2}\;,
\end{equation}
where the free parameters take on the values presented in Table \ref{tab1}. 

Such a choice of parametrization for $\alpha_{\rm P}(t)$ is in part specified by the QCD-motivated asymptotic behavior of the Pomeron Regge trajectory \cite{kearney}, 
\begin{equation}
\label{asyP}
\lim_{t\to-\infty}\alpha_{\rm P}(t)=1\,,
\end{equation}
and by the conditions 
\begin{equation}
\label{deriv}
\frac{d^n\alpha(t)}{dt^n}>0\;\;\;\;(n=1,2...\,;\;t<0)
\end{equation}
which originate from the dispersion relations for Regge trajectories \cite{collins} and are expected to be valid for any Reggeon.\footnote{Other analytic properties of 
parametrizations (\ref{pomeron}) in no way deserve serious consideration. These expressions should be treated just as some nonanalytic quantitative approximations (valid 
at low negative $t$ only) to the corresponding true dynamic functions whose analytic structure is still unknown.}
\begin{table}[ht]
\begin{center}
\begin{tabular}{|l|l|}
\hline
\bf Parameter          & \bf Value                   \\
\hline
$\alpha_{\rm P}(0)-1$  & $0.109\pm 0.017$            \\
$\tau_a$               & $(0.535\pm 0.057)$ GeV$^2$  \\
$g_{pp\rm P}(0)$       & $(13.8\pm 2.3)$ GeV         \\
$a_g$                  & $(0.23\pm 0.07)$ GeV$^{-2}$ \\
\hline
\end{tabular}
\end{center}
\vskip -0.2cm
\caption{The parameter values for (\ref{pomeron}) fitted to the nucleon-nucleon elastic scattering data.}
\label{tab1}
\end{table}

\begin{figure}[ht]
\epsfxsize=8.1cm\epsfysize=8.1cm\epsffile{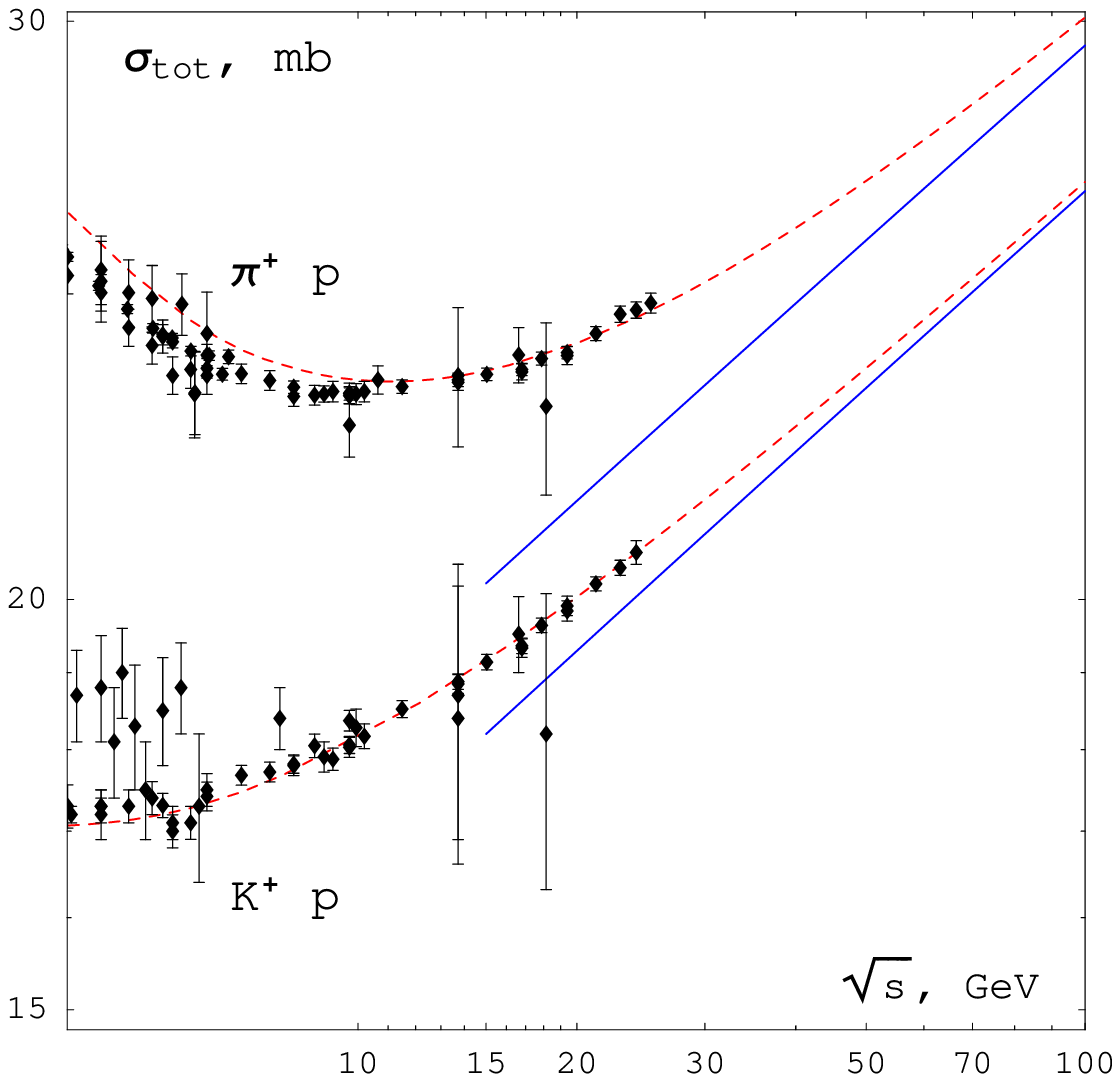}
\vskip -8.25cm
\hskip 8.7cm
\epsfxsize=8.25cm\epsfysize=8.25cm\epsffile{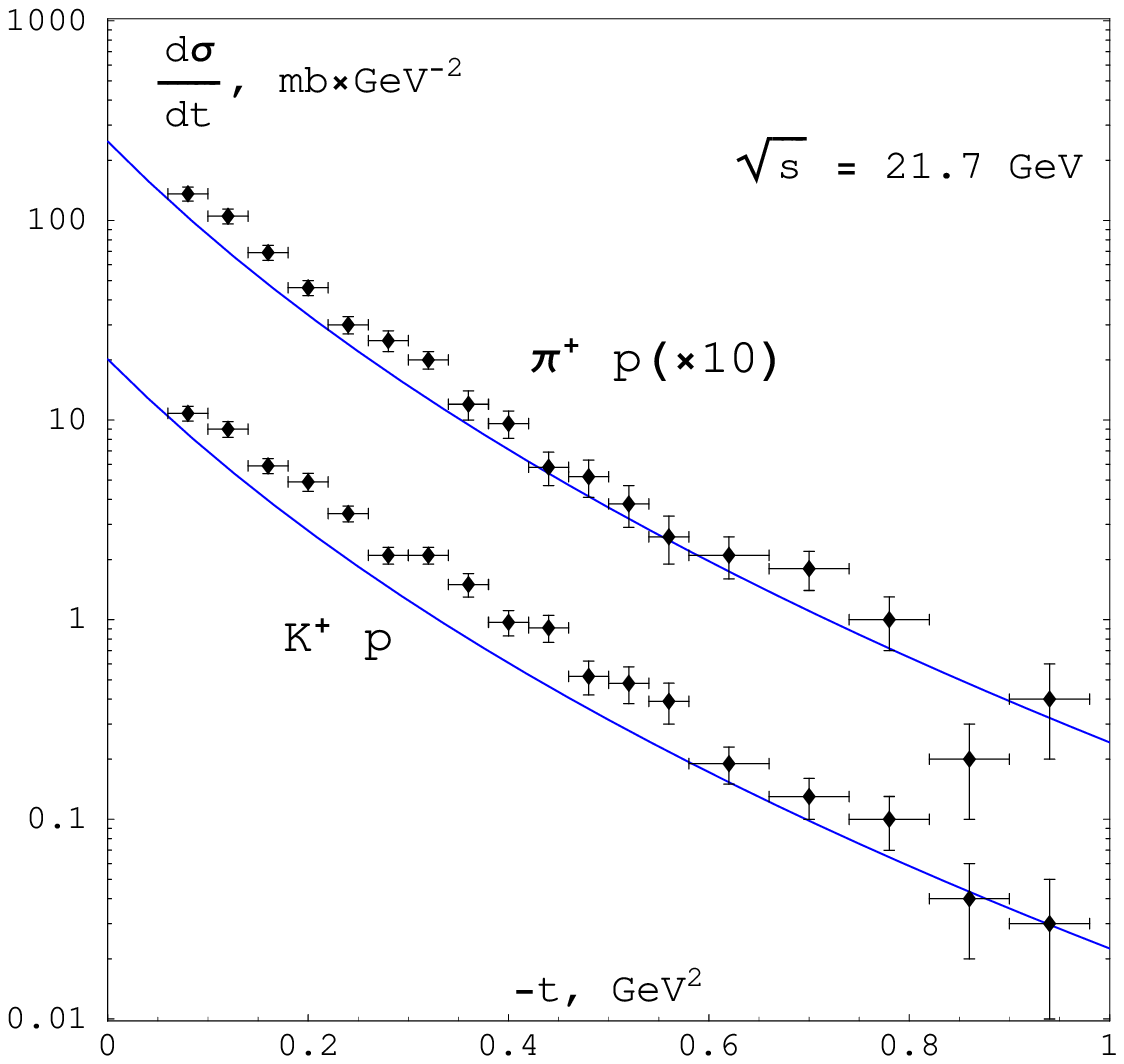}
\caption{The model description of the $\pi^+p$ and $K^+p$ scattering observables \cite{pdg,adamus}. The solid lines correspond to the Pomeron exchange eikonal approximation 
(\ref{eikrepr}),(\ref{pomeron}) of the amplitude. The dashed lines are obtained by adding the secondary Regge pole terms (\ref{dola}) of the DL model for total 
cross-sections \cite{dl}.}
\label{pion}
\end{figure}

Fixing 
\begin{equation}
g_{\pi\pi\rm P}(t) = g_{\pi\pi\rm P}(0) = 8.0\,{\rm GeV}
\label{piPo}
\end{equation}
and
\begin{equation}
g_{KK\rm P}(t) = g_{KK\rm P}(0) = 7.1\,{\rm GeV}\,,
\label{kaPo}
\end{equation}
we come to the description presented in Fig. \ref{pion} (the solid lines). The secondary Reggeon interaction influence on the $\pi^+p$ and $K^+p$ total cross-sections 
can be roughly taken into account via straight addition of the corresponding DL model terms \cite{dl} (see the dashed lines in Fig. \ref{pion}):
\begin{equation}
\Delta\sigma_{tot}^{\pi^+p} = 27.56\left(\frac{s}{s_0}\right)^{-0.4525}{\rm mb}\,,\;\;\;\;\Delta\sigma_{tot}^{K^+p} = 8.15\left(\frac{s}{s_0}\right)^{-0.4525}{\rm mb}\,.
\label{dola}
\end{equation}

The deviation of the model differential cross-section curves from the experimental data cannot be removed so easily, since it requires introduction of extra unknown 
functions of $t$. The main problem is that, in view of the rather high slopes of the secondary Regge trajectories in the resonance region, 
\begin{equation}
\alpha'_{\rm R}(t)\sim 1\,{\rm GeV}^{-2}\;\;\;\;(t>0)\,,
\label{linR}
\end{equation}
their intercept values,
\begin{equation}
\alpha_{\rm R}(0)\sim 0.5\,,
\label{interR}
\end{equation}
their asymptotic behavior in the deeply Euclidean domain \cite{kk}, 
\begin{equation}
\lim_{t\to-\infty}\alpha_{\rm R}(t)=0\,,
\label{asyR}
\end{equation}
and conditions (\ref{deriv}), they are expected to be essentially nonlinear in the diffractive scattering range. Hence, $t$-behavior of the corresponding Regge residues 
turns out strongly related to the nontrivial $t$-behavior of the factors $\alpha'_{\rm R}(t)$ which originate from the Reggeon resonance propagators (for details, see the 
Appendix of \cite{godizov}). As a consequence, reliable determination of the secondary Reggeon exchange contributions into the eikonal (Born amplitude) at nonzero $t$ 
is possible only via combined fitting of all the unknown functions to the overall set of available data on nucleon-nucleon and meson-proton scattering, including 
charge exchange reactions \cite{exchange}. Such a statistically satisfactory description is not achieved yet. It still remains one of the most crucial challenges in Regge 
phenomenology.

Nonetheless, the relative divergence between the presented model curves and the data is limited. The underestimation of $|T(s,t)|$ at $\sqrt{s} = 21.7$ GeV does 
not exceed 16\% and 25\% for the $\pi^+p$ and $K^+p$ scattering, respectively.\footnote{In particular, the model estimation of the $\pi^+p$ differential cross-section 
slope at $t=-0.2$ GeV$^2$,\linebreak $B^{model}\approx 8.7$ GeV$^{-2}$, coincides with the measured value (see Fig. 2 in \cite{adamus}).} This fact might be interpreted as a 
hint to rather weak $t$-dependence of the Pomeron coupling to light scalar mesons in the diffractive scattering regime. Unfortunately, the contamination by secondary Reggeon 
exchanges (though subdominant) does not allow to establish this $t$-behavior explicitly in the framework of approximation (\ref{eikrepr}).\linebreak To elucidate on this 
matter, we need to analyze available data on the high-energy scattering of vector mesons on protons.

\newpage

\subsection*{2.2. Exclusive photoproduction of vector mesons on protons}
\begin{figure}[ht]
\vskip -0.4cm
\epsfxsize=8.4cm\epsfysize=8.4cm\epsffile{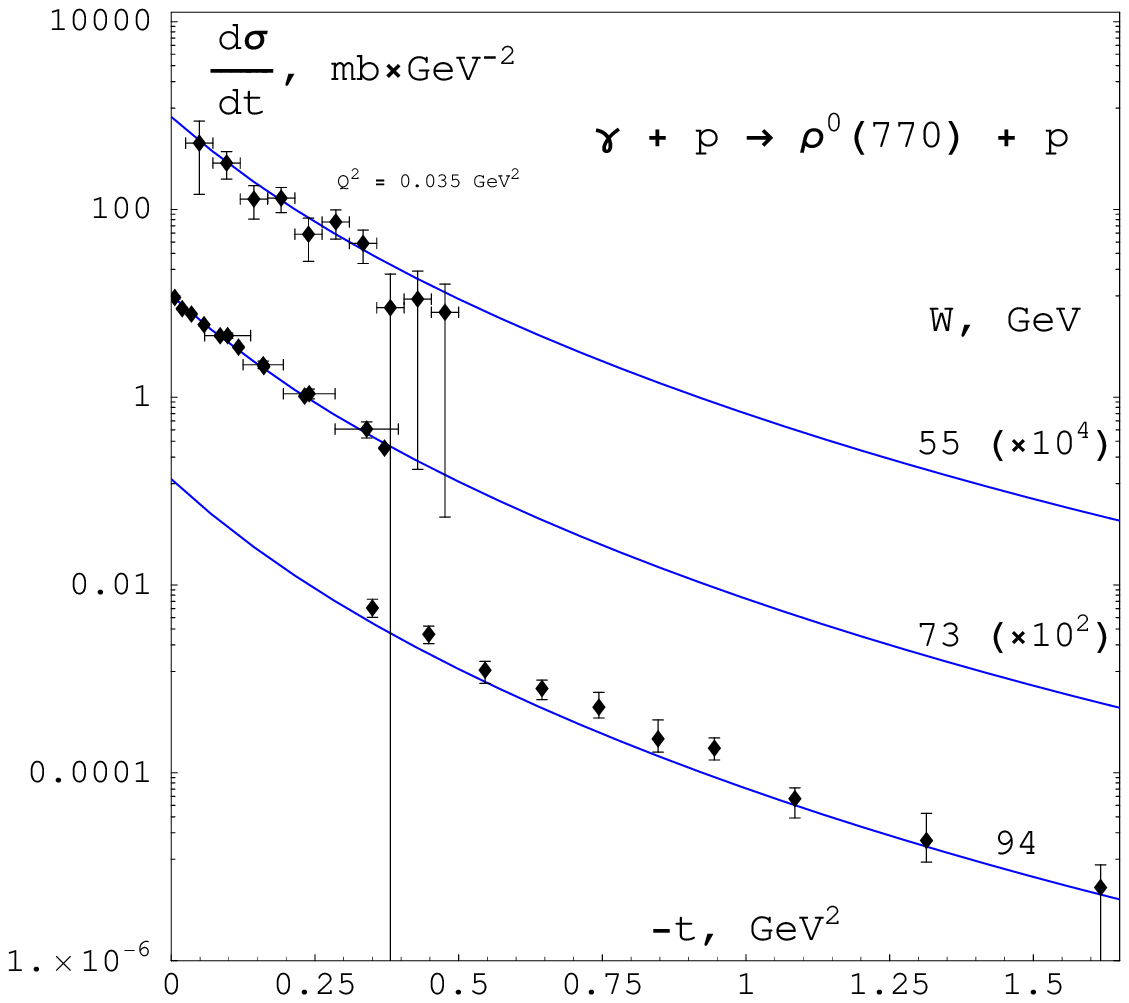}
\vskip -8.3cm
\hskip 8.87cm
\epsfxsize=8.25cm\epsfysize=8.25cm\epsffile{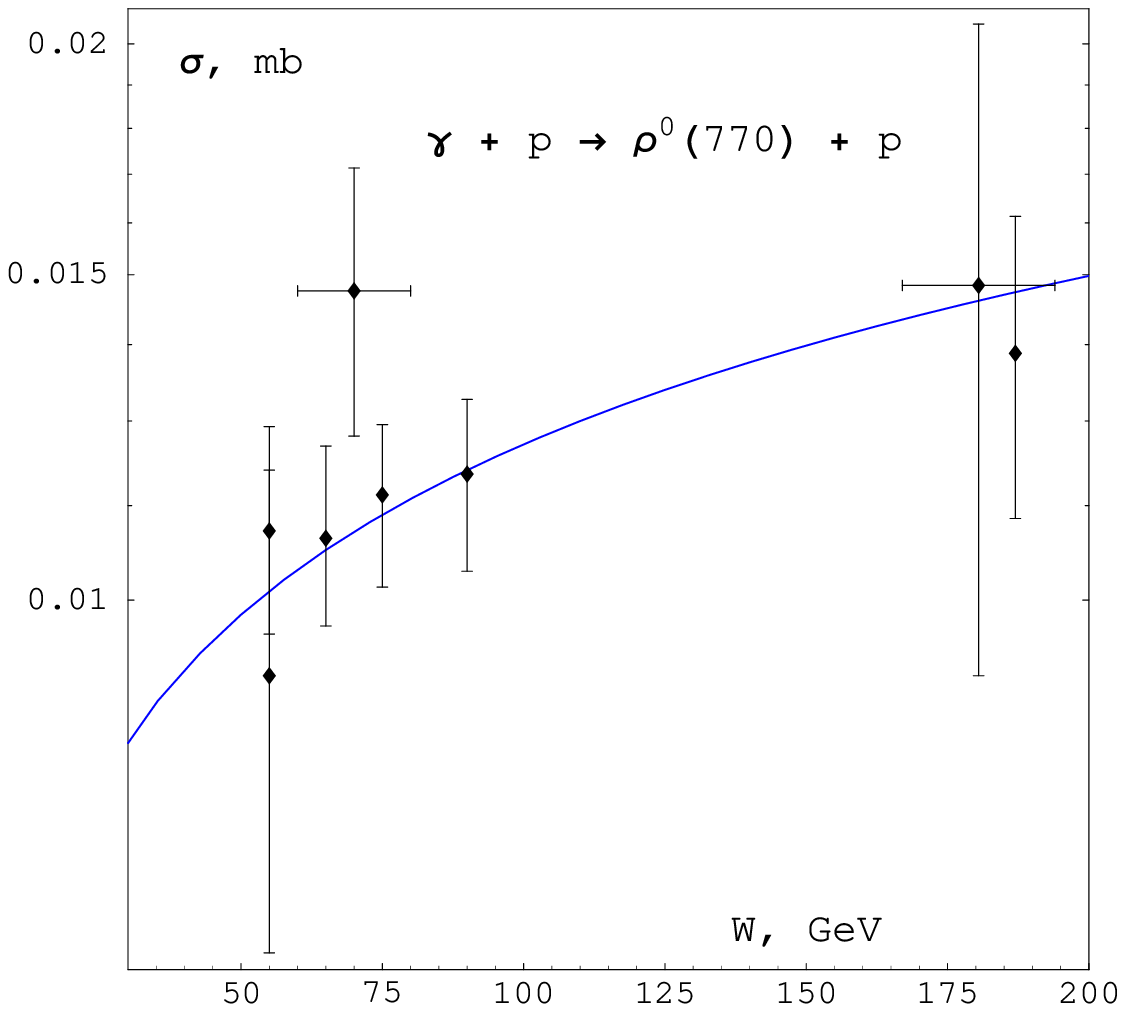}
\vskip -0.3cm
\epsfxsize=8.4cm\epsfysize=8.4cm\epsffile{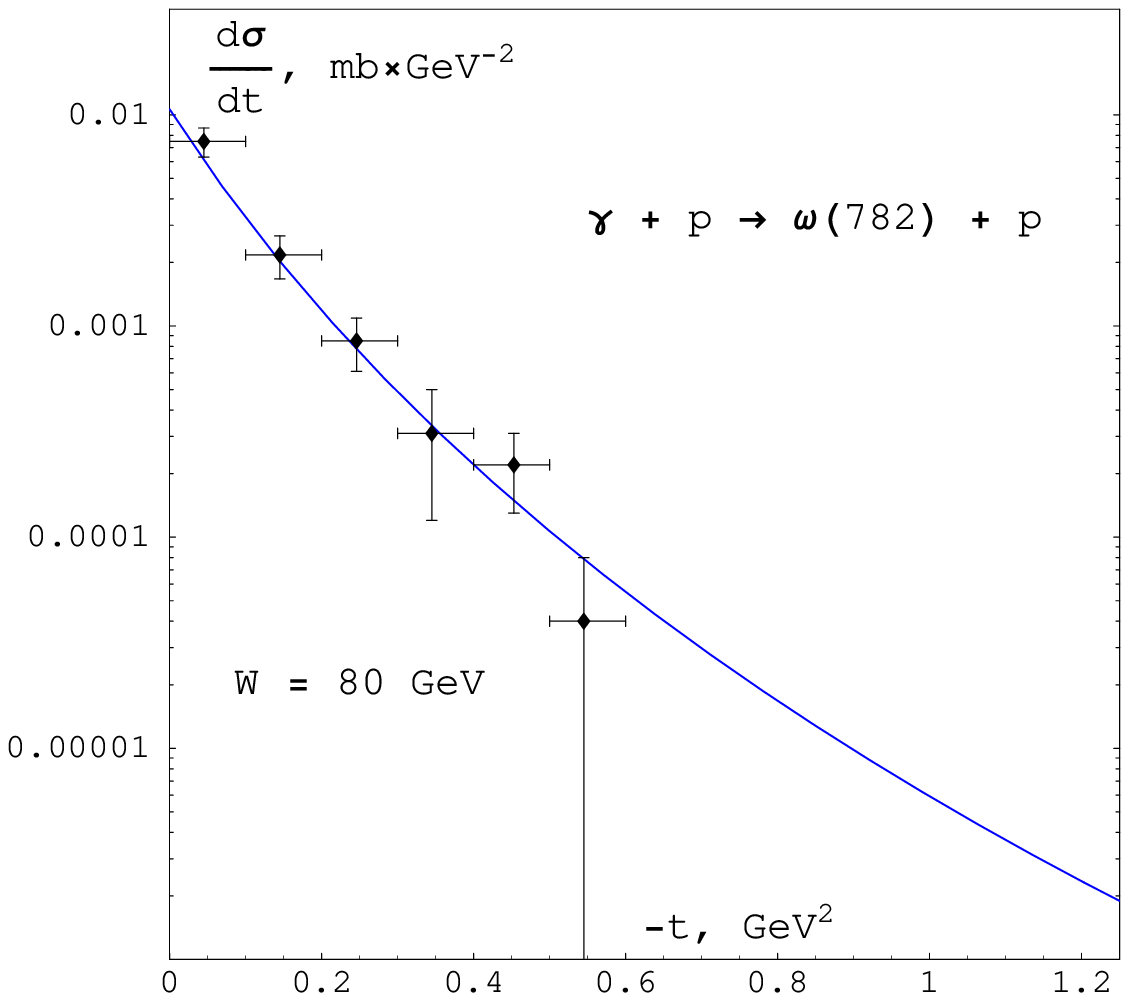}
\vskip -8.4cm
\hskip 8.55cm
\epsfxsize=8.35cm\epsfysize=8.35cm\epsffile{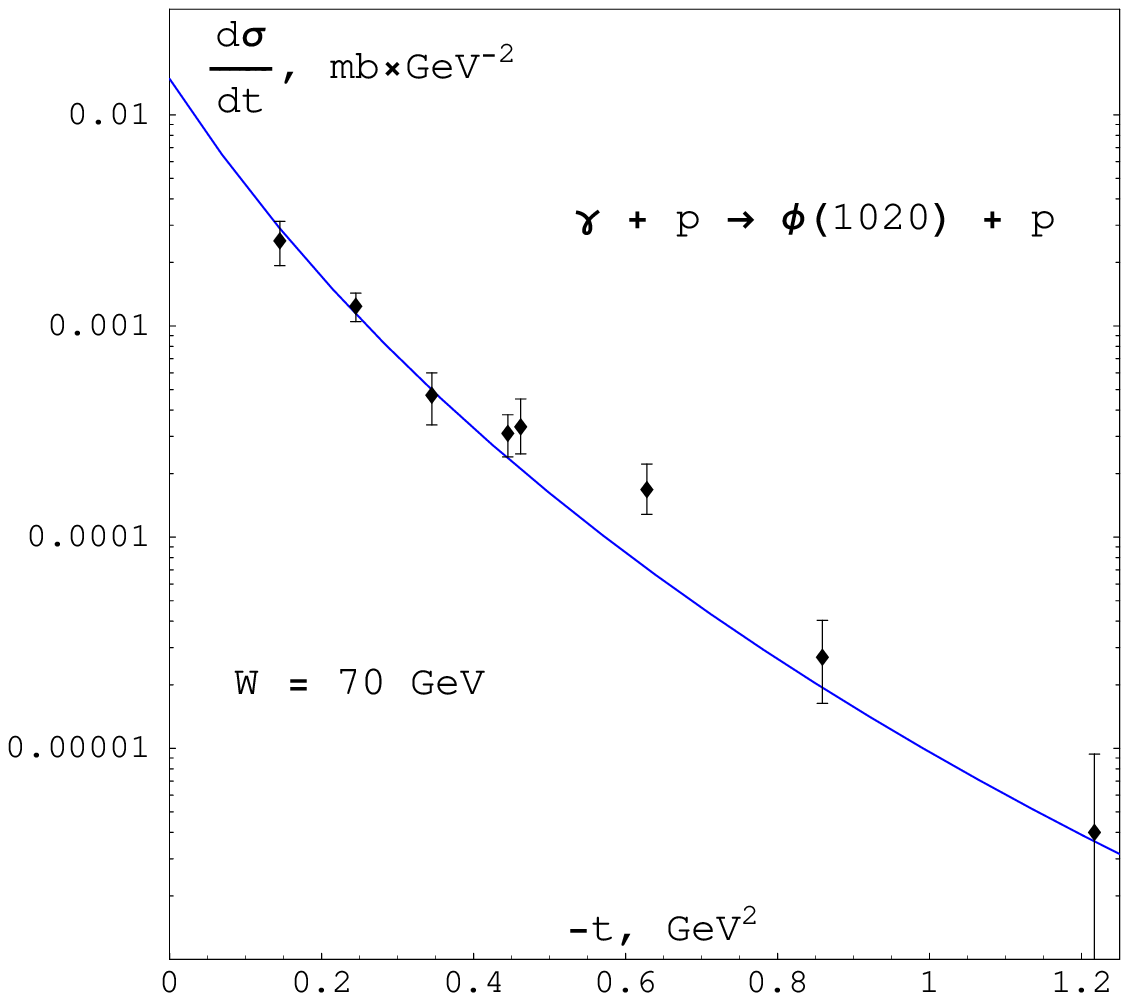}
\vskip -0.4cm
\caption{The model description of the vector meson photoproduction observables \cite{rho,rho1,rho2,rho3,rhophi,omega,phi}.}
\label{vec}
\end{figure}

The vector meson photoproduction data \cite{rho,rho1,rho2,rho3,rhophi,omega,phi} are available in the kinematic range wherein the influence of secondary Reggeon exchanges 
is negligible. According to the hypothesis of $s$-channel helicity conservation (SCHC) confirmed experimentally \cite{helicity}, the cross-sections of reactions 
$\gamma+p\to V+p$ are dominated by the non-flip helicity amplitudes. In the leading approximation, these amplitudes can be represented as \cite{sakurai,petrov} 
\begin{equation}
T_{\gamma p\to V p}(W^2,t)\approx\sum_{V'}C_T^{V'}(0)\,T_{V' p\to V p}(W^2,t)\,,
\label{vecdm}
\end{equation}
where $W$ is the collision energy and $C_T^{V}(0)$ are the vector dominance model (VDM) coefficients: 
\begin{equation}
C_T^{V}(Q^2)=\sqrt{\frac{3\,\Gamma_{V\to e^+e^-}}{\alpha_e M_V}}\frac{M^2_V}{M^2_V+Q^2}
\label{vco}
\end{equation}
($M_V$ is the vector meson mass and $Q^2\ll M_V^2$ is the photon virtuality). It is known that at small values of $t$ the Pomeron diagonal coupling to hadrons is much 
stronger than the off-diagonal one. Otherwise, say, the cross-section of the low-mass diffractive excitation of proton in high-energy $pp$ collisions would be comparable to 
the elastic cross-section. Thus, in the range $W>30$ GeV, 
\begin{equation}
T_{\gamma p\to V p}(W^2,t)\approx C_T^{V}(0)\,T_{V p\to V p}(W^2,t)\,,
\label{vdm}
\end{equation}
where $T_{V p\to V p}(W^2,t)$ is calculated with the help of (\ref{eikrepr}).

Fixing 
\begin{equation}
g_{\omega\omega\rm P}(t) = g_{\rho\rho\rm P}(t) = g_{\rho\rho\rm P}(0) = 7.07\,{\rm GeV}\,,\;\;\;\;g_{\phi\phi\rm P}(t) = g_{\phi\phi\rm P}(0) = 6.7\,{\rm GeV}\,,
\label{vePo}
\end{equation}
we come to the 
description presented in Fig. \ref{vec} and Table \ref{tab2}. The predicted exponential slope of $d\sigma/dt$ for $\gamma+p\to \rho^0(770)+p$ at $W=73$ GeV and 
0.073 GeV$^2<-t<0.4$ GeV$^2$,\linebreak $B^{model}\approx 9.1$ GeV$^{-2}$, is compatible with the measured value, $B^{exp}= (9.8\pm 1.36)$ GeV$^{-2}$ \cite{rho2}. 

The deviation of the $\rho^0(770)$ and $\phi(1020)$ photoproduction curves from the data at\linebreak $-t>0.4$ GeV$^2$ could be explained by the 15\% normalization 
uncertainty \cite{rhophi} not included into the error bars. If we multiply the $\rho^0(770)$ photoproduction data set \cite{rhophi} by factor 0.85 and exclude two outlying 
points, at $t=-0.019$ GeV$^2$ and $t=-0.371$ GeV$^2$, from data set \cite{rho3}, the description quality becomes much better: $\chi^2/N_{DoF}\approx 1.17$ (37 points, 1 free 
parameter).

Fitting the $t$-dependent test expressions of the $\rho^0(770)$-Pomeron coupling,
\begin{equation}
g_{\rho\rho\rm P}(t)=\frac{g_{\rho\rho\rm P}(0)}{1-b\,t}
\label{hype}
\end{equation}
or
\begin{equation}
g_{\rho\rho\rm P}(t)=g_{\rho\rho\rm P}(0)\,e^{b\,t}\,,
\label{eksp}
\end{equation}
we reach the minimum of $\chi^2$ at $g_{\rho\rho\rm P}(0)=7.08$ GeV and $b=0.020$ GeV$^{-2}$ for both the parametrizations. However, the description quality becomes even 
worse, since the slight decrease of $\chi^2$ is overcompensated by the decrease of $N_{DoF}$ due to introduction of extra parameter. Such a result can be interpreted as a 
manifestation of extremely weak $t$-dependence of $g_{\rho\rho\rm P}(t)$ and $g_{\omega\omega\rm P}(t)$ in the diffraction domain.
\begin{table}[ht]
\begin{center}
\begin{tabular}{|l|l|l|l|}
\hline
\bf Data set                  & $W$, GeV     & \bf Number of points &  $\chi_{[i]}^2$   \\
\hline
\cite{rho1}   ($\rho$)        & 55        & 10       &   6.5        \\
\cite{rho2}   ($\rho$)        & 73        & 4        &   7.9        \\
\cite{rho3}   ($\rho$)        & 73        & 9        &  36.4        \\
\cite{rhophi} ($\rho$)        & 94        & 10       &  19.0        \\
\cite{omega}  ($\omega$)      & 80        & 6        &   4.0        \\
\cite{phi}    ($\phi$)        & 70        & 4        &   1.8        \\
\cite{rhophi} ($\phi$)        & 94        & 4        &   8.6        \\
\hline
\end{tabular}
\end{center}
\vskip -0.2cm
\caption{The quality of description of the experimental angular distributions in the vector meson exclusive photoproduction on protons 
at $g_{\omega\omega\rm P}(t) = g_{\rho\rho\rm P}(t) = g_{\rho\rho\rm P}(0) = 7.07$ GeV and $g_{\phi\phi\rm P}(t) = g_{\phi\phi\rm P}(0) = 6.7$ GeV.}
\label{tab2}
\end{table}

In view of the fact that the Pomeron coupling to vector mesons is ``spin-blind'' (SCHC) and comparable to $g_{\pi\pi\rm P}(0)$ and $g_{KK\rm P}(0)$, we could expect the 
existence of some likeness between the Pomeron interactions of vector and scalar mesons and, so, expect a similar weakness of the\linebreak $t$-dependence of the Pomeron 
coupling to light scalar mesons, though this expectation should be considered as just an assumption, until final solution of the above-mentioned secondary Reggeon problem 
is achieved.

The obtained values of $g_{\rho\rho\rm P}(0) = g_{\omega\omega\rm P}(0)$ and $g_{\phi\phi\rm P}(0)$ allow to estimate the Pomeron coupling to real photons.

\subsection*{2.3. High-energy $\gamma p$ scattering}

\begin{figure}[ht]
\begin{center}
\epsfxsize=8.1cm\epsfysize=8.1cm\epsffile{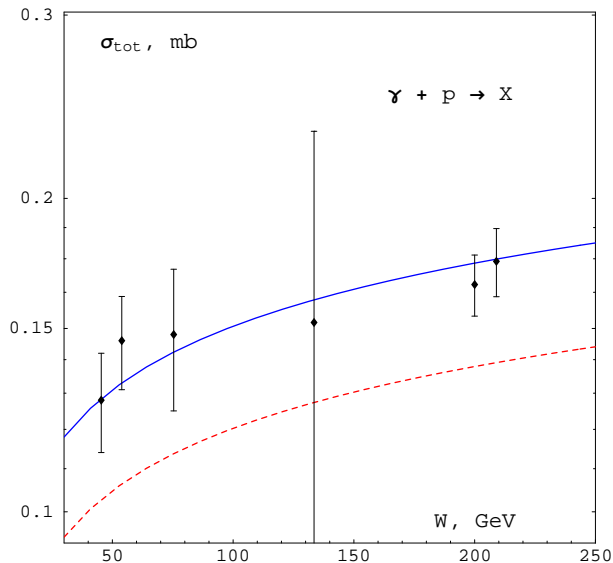}
\end{center}
\vskip -0.7cm
\caption{The total cross-sections of $\gamma p$ scattering \cite{pdg}. The dashed line is the contribution of the fluctuations 
$\gamma\to \rho^0(770),\omega(782),\phi(1020)\to\gamma$.}
\vskip 0.3cm
\label{gamma}
\end{figure}

Owing to the optical theorem: $W^2\sigma^{\gamma p}_{tot}(W)={\rm Im}\,T_{\gamma p\to\gamma p}(W^2,0)$. In the range $W>30$ GeV, the diagonal fluctuations of photons to 
mesons $\rho^0(770)$, $\omega(782)$, and $\phi(1020)$  provide just about 80\% of $\sigma^{\gamma p}_{tot}$ (see the dashed line in Fig. \ref{gamma}). Other 
contributions can be roughly taken into account via introduction of the pole-like term 
$A(W)=g_A(0)\,g_{pp\rm P}(0)\,\pi\alpha'_{\rm P}(0)\left(\frac{W^2}{2s_0}\right)^{\alpha_{\rm P}(0)}$, where $g_A(0)=0.007$ GeV, so that the full amplitude is 
approximated as (the solid line in Fig. \ref{gamma})
\begin{equation}
{\rm Im}\;T_{\gamma p\to\gamma p}(W^2,0)\;\;\approx \sum_{V=\rho,\omega,\phi}[C_T^{V}(0)]^2\,{\rm Im}\,T_{V p\to V p}(W^2,0)\;\;+\;\;A(W)\,.
\label{gamfull}
\end{equation}
The total contribution of absorptive corrections to the imaginary part of the $\rho^0p$ scattering forward amplitude is negative. In the interval 40 GeV$<W<210$ GeV, it  
does not exceed 16\% of the corresponding Born amplitude. Hence, we do not expect that the underestimation of the Pomeron coupling to photon which emerges under 
fitting $g_A(0)$ to the data is higher than 20\% of this parameter, since the Pomeron interaction of the heavier vector mesons is not expected to be stronger than 
of $\rho^0(770)$ meson. So and thus, we obtain the following estimation of $g_{\gamma\gamma\rm P}(0)$: 
\begin{equation}
g_{\gamma\gamma\rm P}(0)=\sum_{V=\rho,\omega,\phi}[C_T^{V}(0)]^2\,g_{VV\rm P}(0)+\kappa\,g_A(0)\;\;\;\;\left(1<\kappa<1.2\right)\,.
\label{gaspace}
\end{equation}

\section*{3. The Pomeron decays to light mesons and photons}

Reggeon coupling to light mesons in the diffraction and resonance domains is driven by the same structure functions (see formulae (\rm A.3) and (\rm A.4) of the 
Appendix). Although these formulae are initially derived for spinless particles, they can be adapted to vector mesons and photons. 

Let us formulate {\it the comparability assumption} which underlies the further discussion:
\begin{itemize}
\item The Pomeron coupling to light mesons in the resonance region is comparable or even approximately equal to its coupling to the corresponding particles in the diffractive 
scattering regime. In particular, $|g_{hh{\rm P}}(2;\,M_{\rm P}^2,m_h^2,m_h^2)|\sim |g_{hh{\rm P}}(0)|\equiv g_{hh\rm P}(\alpha_{\rm P}(0);\,0,m_h^2,m_h^2)$, where 
$M_{\rm P}$ is the Pomeron ground state mass.
\end{itemize} 
 
Though this assumption is highly nontrivial (due to the high sensitivity of the resonance partial decay width to the value of its coupling to the decay products), 
it does not seem to be exceptional. For example, as applied to $\rho$-mesons \cite{pdgf2}, formula (\rm A.4) yields 
\begin{equation}
\frac{|g_{\pi\pi\rho}(3;\,M_{\rho_3(1690)}^2,m_\pi^2,m_\pi^2)|}{|g_{\pi\pi\rho}(1;\,M_{\rho(770)}^2,m_\pi^2,m_\pi^2)|}=1.1\pm 0.05\,.
\label{rho}
\end{equation}
From the physical standpoint, such a weak $t$-dependence just implies that the corresponding Reggeon perceives pions (and, possibly, some other mesons) as pointlike 
particles. 

Regarding the Pomeron spin-2 resonance decays to pairs of light vector mesons or photons, the comparability assumption supplemented by the above-mentioned SCHC hypothesis 
(which is valid in the diffraction region) implies the dominance of the only orbital mode: $L=0$. Other orbital modes are related to helicity amplitudes conjugated to the 
spin-flip vertices in Pomeron exchanges. Hence, they are expected to be dynamically supressed on the same reasons as the spin-flip amplitudes in the 
high-energy diffractive scattering. Consequently, in the case of Pomeron interaction, formula (\rm A.4) is applicable, as well, to the decays to pairs of vector mesons or 
photons (of course, the spin-2 resonance decays to pairs of scalar mesons take place in the $L=2$ orbital mode only).

Let us try to describe the $f_2(1950)$ resonance observables with the help of the above-stated assumption. 
For example, the constant related to the Pomeron ground state decay to two real photons of different polarization can be estimated as 
\begin{equation}
|g_{\gamma\gamma{\rm P}}(2;\,M^2_{\rm P},0,0)|\;=\;g_{\gamma\gamma{\rm P}}(0)=(39.7\pm 0.7)\,{\rm MeV}\,. 
\label{gaga}
\end{equation}
Inserting $M_{\rm P} = M_{f_2(1950)} = (1.944\,\pm\,0.012)$ GeV \cite{pdgf2} into (\rm A.4), we come to the following result: 
\begin{equation}
\Gamma_{f_2(1950)\to \gamma\gamma}=(960\pm 50)\,{\rm eV}\,. 
\label{Pgaga}
\end{equation}
In addition, with the help of estimations (\ref{piPo}) and (\ref{kaPo}), we predict 
\begin{equation}
\Gamma_{f_2(1950)\to\pi^0\pi^0}=(37\pm 1)\,{\rm MeV}\,,\;\;\;\;\Gamma_{f_2(1950)\to K^+K^-}=(29\pm 1)\,{\rm MeV}\,, 
\label{meme}
\end{equation}
and, then, in view of $\Gamma_{f_2(1950)\to X}=(472\pm 18)$ MeV \cite{pdgf2}, obtain 
\begin{equation}
\frac{\Gamma_{f_2(1950)\to\gamma\gamma}\Gamma_{f_2(1950)\to\pi^0\pi^0}}{\Gamma_{f_2(1950)\to X}}=(75\,\pm\,5)\;{\rm eV}\,,\;\;\;\;\;\;
\frac{\Gamma_{f_2(1950)\to\gamma\gamma}\Gamma_{f_2(1950)\to K^+K^-}}{\Gamma_{f_2(1950)\to X}}=(59\,\pm\,4)\;{\rm eV}\,,
\label{piondecay}
\end{equation}
while the corresponding experimental values are \cite{pion,kaon}
\begin{equation}
\frac{\Gamma_{f_2(1950)\to\gamma\gamma}\Gamma_{f_2(1950)\to\pi^0\pi^0}}{\Gamma_{f_2(1950)\to X}}=54^{+23}_{-14}\;{\rm eV}\,,\;\;\;\;\;\;
\frac{\Gamma_{f_2(1950)\to\gamma\gamma}\Gamma_{f_2(1950)\to K^+K^-}}{\Gamma_{f_2(1950)\to X}}=(61\pm 2\pm 13)\;{\rm eV}.
\label{belledecay}
\end{equation}

The same procedure applied to $f_2(2300)$ resonance ($M_{f_2(2300)} = (2.297\,\pm\,0.028)$ GeV, $\Gamma_{f_2(2300)\to X}=(149\pm 40)$ MeV \cite{pdgf2}) yields the 
estimation 
\begin{equation}
\frac{\Gamma_{f_2(2300)\to\gamma\gamma}\Gamma_{f_2(2300)\to K\bar K}}{\Gamma_{f_2(2300)\to X}}=(1300\,\pm\,370)\;{\rm eV}\,,
\label{f2300}
\end{equation}
which catastrophically diverges from the Belle data \cite{belle13}:
\begin{equation}
\frac{\Gamma_{f_2(2300)\to\gamma\gamma}\Gamma_{f_2(2300)\to K\bar K}}{\Gamma_{f_2(2300)\to X}}=3.2^{+0.5+1.3}_{-0.4-2.2}\;{\rm eV}.
\label{belle2300}
\end{equation}
This inconsistency allows to exclude $f_2(2300)$ resonance from the candidates for the Pomeron ground state. Unfortunately, analogous data for 
$f_2(2010)$ and $f_2(2340)$ resonances \cite{pdgf2} are currently unavailable.

\section*{4. Discussion}

So and thus, direct propagation of the Pomeron couplings $g_{KK{\rm P}}$ and $g_{VV{\rm P}}$ from the diffraction to resonance region yields a correct estimation of 
$\Gamma_{f_2(1950)\to\gamma\gamma}\,\Gamma_{f_2(1950)\to K^+K^-}/\Gamma_{f_2(1950)\to X}$. 

The model and experimental estimations of $\Gamma_{f_2(1950)\to\gamma\gamma}\,\Gamma_{f_2(1950)\to \pi^0\pi^0}/\Gamma_{f_2(1950)\to X}$ overlap as well, but the significant 
systematic uncertainties \cite{pion} not included into the presented value (\ref{belledecay}) make this result not so impressive.

\begin{figure}[ht]
\begin{center}
\epsfxsize=8.1cm\epsfysize=8.1cm\epsffile{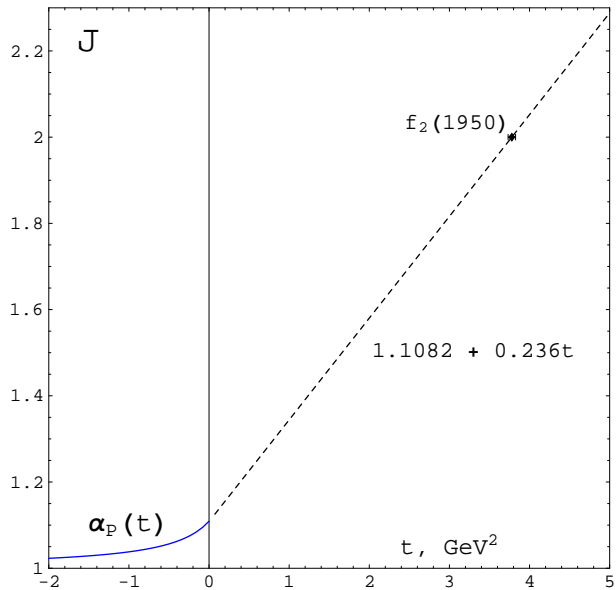}
\end{center}
\caption{Expected behavior of the Chew-Frautschi plot for the Pomeron. The solid line is the Pomeron Regge trajectory (\ref{pomeron}) fitted to the nucleon-nucleon 
scattering data. The dashed line corresponds to $\alpha^{lin}_{\rm P}(t)=1.1082+0.236\,t$.}
\label{tra}
\end{figure}

A quite independent argument in favor of our reasoning is the $f_2(1950)$ resonance location in the $t-J$ plane (see Fig. \ref{tra}), which allows a smooth linkage 
between the linear Chew-Frautschi plot and the nonlinear Regge trajectory of the Pomeron (\ref{pomeron}) extracted from the nucleon-nucleon scattering data. 
This coincidence encourages to consider more seriously the possibility of the $f_2(1950)$ resonance correspondence to the Pomeron ground state, since other observed 
$0^+(2^{++})$ resonances do not seem to be so good candidates.

Of course, at present, the discussed coincidences should be qualified just as some noteworthy indications. For a firm conclusion, we need new and, if possible, more 
accurate data on both the $f_2(1950)$ resonance observables and the $\pi p$ and $K p$ scattering at higher energies. Another prediction which should be 
checked is the opposite polarization of the photons in $f_2(1950)\to\gamma\gamma$ decays which follows from the predicted above dominance of the $L=0$ orbital mode.

In fact, the possible physical consequences of the presumed comparability are much deeper than just a practical feasibility of the Pomeron 
identification. In the Euclidean domain, the behavior like \{$\alpha_{\rm P}(t)>1$, $\lim_{t\to -\infty}\alpha_{\rm P}(t)=1$\} is characteristic for glueball Reggeons (for 
details, see \cite{kearney,low,heckathorn,wu,kirschner,petrov2}). If the assumption is correct, then the Pomeron is expected to keep its glueball nature in the resonance 
region as well. In other words, if the predictions, regarding the $K\bar K$, $\pi\pi$, and $\gamma\gamma$ decays, are confirmed, we would have enough grounds 
{\it to consider $f_2(1950)$ resonance as not only the Pomeron ground state, but the lightest tensor glueball}. 

The idea that $f_2(1950)$ resonance as a glueball is, certainly, not new. Although its mass is significantly lower than the predictions by lattice QCD \cite{lat}, 
it is discussed as a glueball candidate from time to time, as because of the high value of its decay width \cite{holo}, so due to some features of the high-energy 
$p+p\to p+f_2(1950)+p$ differential cross-sections \cite{wa102}.

\subsection*{Acknowledgements} The author is indebted to his colleagues from the IHEP Division of Theoretical Physics for helpful discussions.

\section*{Appendix. Interrelation between Reggeon dynamics in the diffraction and resonance regions}

Let us consider the elastic interaction of two scalar mesons via exchange by single virtual spin-$j$ particle. The contribution of such exchange to the Born amplitude can be 
represented as 
$$
\delta^{(j,m_j)}(p_1,p_2,\Delta) = J^{(1)}_{\alpha_1...\alpha_j}(p_1,\Delta)\,\frac{D_{(M_j)}^{\alpha_1...\alpha_j,\beta_1...\beta_j}(\Delta)}{m_j^2-\Delta^2}\;
J^{(2)}_{\beta_1...\beta_j}(p_2,-\Delta)\;,
\eqno{(\rm A.1)}
$$
where $\frac{D_{(M_j)}^{\alpha_1...\alpha_j,\beta_1...\beta_j}(\Delta)}{m_j^2-\Delta^2}$ is the spin-$j$ particle propagator, $m_j^2=M_j^2-iM_j\Gamma_j$ ($M_j$ and 
$\Gamma_j$ are the mass and decay width of the corresponding resonance), $J^{(1,2)}_{\alpha_1...\alpha_j}$ are the currents of the scalar mesons, $p_1$ and $p_2$ are 
the 4-momenta of the incoming particles, and $\Delta$ is the transferred 4-momentum. 

Obviously, in the kinematic range $s\equiv (p_1+p_2)^2\gg \{|p_{1,2}^2|,|\Delta^2|,|(p_{1,2}\Delta)|\}$ (\rm A.1) transforms into 
$$
\delta^{(j,m_j)}(p_1,p_2,\Delta) = \frac{g^{(1)}(j;\,p_1^2,\Delta^2,(p_1\Delta))\;g^{(2)}(j;\,p_2^2,\Delta^2,-(p_2\Delta))}{m_j^2-\Delta^2}\left(\frac{s}{2s_0}\right)^j
\left[1+O\left(\frac{1}{s}\right)\right]\,,
\eqno{(\rm A.2)}
$$
where $g^{(1,2)}$ are the structure functions at the tensor structure $\frac{p_{\alpha_1}...p_{\alpha_j}}{s_0^{j/2}}$ in the currents $J^{(1,2)}$ and $s_0\equiv 1$ GeV$^2$ 
is a scale (unit of measurement) characteristic for hadron physics. 

The last expression can be used for derivation of the appropriate Reggeon exchange contribution to the Born amplitude for high-energy elastic diffractive 
scattering (see \cite{collins} or, say, the Appendix of \cite{godizov}): 
$$
\delta_{\rm R}(s,t) = g_{\rm R}^{(1)}(\alpha_{\rm R}(t);\,t,m_1^2,m_1^2)\;g_{\rm R}^{(2)}(\alpha_{\rm R}(t);\,t,m_2^2,m_2^2)\;\xi(\alpha_{\rm R}(t))\,
\pi\alpha'_{\rm R}(t)\left(\frac{s}{2s_0}\right)^{\alpha_{\rm R}(t)},
\eqno{(\rm A.3)}
$$
where $t\equiv\Delta^2<0$, the function $\alpha_{\rm R}(t)$ is the corresponding Regge trajectory, and $\xi(\alpha)$ is the so-called Reggeon signature factor 
($\xi(\alpha)=i+{\rm tan}\,\frac{\pi(\alpha-1)}{2}$ for even Reggeons and $\xi(\alpha)=i-{\rm cot}\,\frac{\pi(\alpha-1)}{2}$ for odd Reggeons).

To consider the two-scalar-meson decay of a spin-$j$ resonance state associated with some Reggeon we need to continue (\rm A.1) analytically outside of the $s$-channel 
physical range. If $p_1=-p_2$, $p_1^2=p_2^2=m_a^2$, $(p_1-\Delta)^2=(p_2+\Delta)^2=m_b^2$, and $\Delta^2\to M_j^2>(m_a+m_b)^2$, then, in view of the general properties 
of $D_{(M_j)}^{\alpha_1...\alpha_j,\beta_1...\beta_j}(\Delta)$ at $\Delta^2 = M_j^2$ (the transversality with respect to $\Delta_{\alpha_k}$ ($k = 1,...,j$) and the 
tracelessness with respect to any pair of Lorentz indices like $\{\alpha_i\alpha_j\}$ and $\{\beta_i\beta_j\}$), we obtain the following expression for the resonance 
partial decay width:
$$
\Gamma^{(j)}_{{\rm R}\,\to\,a\,+\,b} = \frac{\lambda^{1/2}(M_j^2,m_a^2,m_b^2)}{16\,\pi\,M_j^3}\;|A^{(j)}_{{\rm R}\,\to\,a\,+\,b}|^2 = 
$$
$$
= \frac{\lambda^{1/2}(M_j^2,m_a^2,m_b^2)}{16\,\pi\,M_j^3}\;\frac{|g_{ab\rm R}(j;\,M_j^2,m_a^2,m_b^2)|^2}{(2j+1)\;s_0^j}\;
p_{\alpha_1}...p_{\alpha_j}\,p_{\beta_1}...p_{\beta_j}\,D_{(M_j)}^{\alpha_1...\alpha_j,\beta_1...\beta_j}(\Delta) = 
\eqno{(\rm A.4)}
$$
$$
= \frac{(j!)^2\,\lambda^{j+1/2}(M_j^2,m_a^2,m_b^2)}{16\,\pi\,2^j\,(2j+1)!\,M_j^{2j+3}s_0^j}\;|g_{ab\rm R}(j;\,M_j^2,m_a^2,m_b^2)|^2\,,
$$
where $\lambda(x,y,z)\equiv x^2+y^2+z^2-2\,x\,y-2\,y\,z-2\,x\,z$.

\end{document}